\documentclass[useAMS,usenatbib]{mn2e}

\usepackage{epsfig}

\newcommand{\mnras}{{MNRAS}}
\newcommand{\apj}{{ ApJ}}
\newcommand{\prd}{{ PRD}}
\newcommand{\physrep}{{ Physics Reports}}

\newcommand{\Pchi}{{\cal P}_{\chi}(k)}
\newcommand{\zre}{z_{\rm eff}}
\newcommand{\Mpci}{{\rm Mpc}^{-1}}
\newcommand{\nrun}{n_{\rm run}}

\def\ltsima{$\; \buildrel < \over \sim \;$}
\def\simlt{\lower.5ex\hbox{\ltsima}}
\def\gtsima{$\; \buildrel > \over \sim \;$}
\def\simgt{\lower.5ex\hbox{\gtsima}}

\newcommand\Url[1]{${\rm #1}$}
\title
[Reconstructing the primordial power spectrum]
{Reconstructing the primordial power spectrum}
\author[S.L.~Bridle et al.]
{S.L.~Bridle$^1$\thanks{E-mail: sarah@ast.cam.ac.uk},
A.M.~Lewis$^2$, J.~Weller$^1$ and G.~Efstathiou$^1$\\
$^1$Institute of Astronomy, University of Cambridge, Madingley Road,
    Cambridge CB3 0HA, UK\\
$^2$CITA, 60 St. George St, Toronto M5S 3H8, ON, Canada\\
}

\date{\today}
\pagerange{\pageref{firstpage}--\pageref{lastpage}}
\pubyear{2003}
\begin{document}

\maketitle

\label{firstpage}

\begin{abstract}
We reconstruct the shape of the primordial power spectrum
from the latest cosmic microwave background data, including the new
results from the Wilkinson Microwave Anisotropy Probe (WMAP),
and large scale structure data from the two degree field galaxy redshift 
survey (2dFGRS). We tested four parameterizations taking into 
account the uncertainties in four cosmological
parameters. First we parameterize the initial spectrum by a tilt and
a running spectral index, finding 
marginal evidence for a running spectral index 
only if the first three WMAP multipoles ($\ell=2,3,4$) 
are included in the analysis. Secondly,
to investigate further the low CMB large scale power, we modify the
conventional power-law spectrum by introducing a scale above
which there is no power. We find a preferred position
of the cut at $k_c\sim 3\times 10^{-4} {\rm Mpc}^{-1}$
although $k_c=0$ (no cut) is not ruled out. 
Thirdly we use a model independent parameterization,
with 16 bands in wavenumber, and find no obvious
sign of deviation from a power law spectrum 
on the scales investigated. 
Furthermore the values of the other cosmological parameters
defining the model remain relatively well constrained despite the
freedom in the shape of the initial power spectrum.
Finally we investigate a model motivated by double inflation, 
in which the 
power spectrum has a break between two characteristic wavenumbers.
We find that if a break is required to be in the range 
$0.01 <k /{\rm Mpc^{-1}} <0.1$ 
then the ratio of amplitudes across the break is constrained to be
$1.23 \pm 0.14$. Our results are consistent with a 
power law spectrum that is featureless and close to scale invariant 
over the wavenumber range $0.005 \simlt k / {\rm Mpc^{-1}} \simlt
0.15$, with a hint of a decrease in power on the largest scales.
\end{abstract}


\section{Introduction}

Measurements of the Cosmic Microwave Background (CMB) radiation
have taken a leap forward with the recent announcement
of the findings of the Wilkinson Microwave Anisotropy Probe (WMAP).
With these new data it is possible to set important constraints
on the shape of the primordial power spectrum and hence to begin
to differentiate between the plethora of models for the early universe.
One of the most intriguing  results comes from a 
a combined analysis of the WMAP data with the two degree
field galaxy redshift survey (2dFGRS) data
\citep{Spergel03,Peiris03},
which indicates that the primordial power spectrum 
might have curvature. The addition of Lyman-$\alpha$ forest data on
smaller scales to strengthen 
this conclusion is however contentious~\citep{Seljak03}.
The low quadrupole
and octopole observed in the CMB temperature power 
spectrum~\citep{Spergel03}
has a low probability in standard models, and may be an indication of
some feature in the initial power spectrum on very large scales.

Both inflationary Big-Bang~\citep{guth81,Linde:82,Albrecht:82} and
more speculative 
cyclic ekpyrotic~\citep{steinhardtt02} models of the early universe
predict very nearly Gaussian
scalar perturbations in the primordial radiation dominated era.
The shape of the perturbation power spectrum depends on
the exact model, which typically involves various unknown
parameters. The objective of this paper is to constrain the shape of
the initial power spectrum directly from observational data with
as few assumptions as possible.

A wealth of cosmological information can be obtained
from analysing the shape of the cosmic microwave background (CMB)
radiation temperature fluctuation power spectrum.
However it has been shown that the effect of changing
the cosmological parameters can be exactly mimicked by
changes in the shape of the primordial power 
spectrum~\citep{souradeepbket98,kinney01}. 
By including measurements of 
the CMB polarization and the late time matter power
spectrum, as measured for example by galaxy redshift surveys, 
the degeneracy can be broken because the cosmological parameters
affect these data in different ways. 
In this paper we combine temperature and
polarization data from the WMAP observations and other CMB
observations on smaller scales with
constraints on the matter power spectrum from the 
2dFGRS~\citep{percivalea02}. 

Since the initial power spectrum is an unknown function one is forced
to parameterize it. There are numerous possibilities.
\cite{wangss99}, \cite{wangm02} and \cite{mukherjeew03c} 
use a number of 
bands in wavenumber. \cite{mukherjeew03a,mukherjeew03b,mukherjeew03c} 
also use a model independent approach but using wavelets.
\cite{barrigagss01} test a particular inflationary model, in which 
a phase transition briefly halts the slow roll of the inflaton.

Inflationary models generically predict a monotonically
slowly varying power spectrum, determined by the shape of the 
inflationary potential.
More complicated inflationary models can give more interesting spectra at the
expense of introducing parameters that are fine tuned to give effects in the
small range of observable wavenumbers.
In this paper we consider various different power spectrum 
parameterizations,
motivated by theoretical models or features of the observed data.
We also adopt a model independent approach which allows
general trends or unexpected features to be detected.

In Section~\ref{framework} we specify the framework
used in the rest of the paper. 
We summarize the results of the conventional power spectrum
parameterization (ie. a power law slope and running spectral
index) in Section~\ref{vanilla},
and discuss the possible implications of the small large scale power
detected by WMAP in Section~\ref{cutoff}.
Section~\ref{pinibins} describes a reconstruction of the 
primordial power spectrum in wavenumber bands on sub-horizon scales.
We test a double inflation model over a particular range of
wavenumbers in Section~\ref{barriga}.
In each section we discuss the implications for
estimates of the other cosmological parameters.

\section{Framework}
\label{framework}

The primordial scalar power spectrum $P_{\chi}$ is defined by
$\langle |\chi|^2 \rangle = \int {{\rm d} \ln k} \, {{\cal P}_{\chi}(k)}$,
where $\chi$ is the
super-horizon comoving curvature perturbation in the early radiation
dominated era.
The commonly assumed power-law power spectrum parameterization is then
\begin{equation}
\label{powerdef}
{{\cal P}_{\chi}(k)} = A_s \left(\frac{k}{k_{s0}}\right)^{n_s-1}. 
\end{equation}
Here $n_s(k) =  d \ln {{\cal P}_{\chi}(k)} / d \ln k\, +1$
is the conventional definition of the scalar 
spectral index, where $n_s=1$ corresponds to a
scale invariant power spectrum
(we use $k_{s0} = 0.05 {\rm Mpc}^{-1}$ throughout). 
The power spectrum amplitude $A_s$
determines the variance of the fluctuations, with $A_s^{1/2} \sim
5\times 10^{-5}$ to give the observed CMB anisotropy amplitude.

In slow roll inflationary models, it is expected that the spectrum
varies only very slowly and that $|n_s-1|$ is much smaller than unity
\citep{Lyth:99}. In general there is a direct relation between the potential
of the inflaton field and the spectral index.
As the potential evolves during inflation the spectral index can vary
slightly as different modes leave the horizon.
This can be characterized by including a second order term in 
the logarithmic expansion of the power spectrum $n_{\rm run}$,
defined by
\begin{equation}
\ln {{\cal P}_{\chi}(k)} = \ln A_s + (n_s -1) \ln 
\left(\frac{k}{k_{s0}} \right)  
+ \frac{n_{\rm run}}{2} \left( \ln \left( \frac{k}{k_{s0}} \right)\right)^2. 
\end{equation}
The value of $n_s$ therefore depends on the pivot scale used, 
for example to convert
to a new pivot scale the relation is
$n_{\rm s}(k_{s0}')=n_{\rm s}(k_{s0}) + n_{\rm run} \log (k_{s0}' /
k_{s0})$. 
More generally double inflation models or multiple field inflation
can lead to breaks and spikes in the primordial power spectrum,
see eg.~\cite{linde90}.
This motivates a more general parameterization of the primordial
power spectrum in terms of amplitudes over discrete bands
in wavenumber.

The primordial power spectrum is related to the linear CMB 
anisotropy power spectrum $C^{XY}_\ell$ by a transfer function 
$T_{\ell}^X (k)$ via
\begin{equation}
C^{XY}_\ell \propto \int {\rm d} \ln k {{\cal P}_{\chi}(k)} T_{\ell}^X(k) T_{\ell}^Y(k).
\end{equation}
where $X$ and $Y$ denote the various temperature and polarization
power spectra.
The transfer function for a mode of wavenumber $k$ peaks at a
multipole of about $\ell \simeq k d_{\rm A}$ where $d_{\rm A}$ is the
angular diameter distance. 
For the concordance model the relation is roughly
$\ell\sim (1e4 {\rm Mpc}) k$.
However the detailed shape of the transfer
function is complicated and a range of wavenumbers contribute to each multipole
(see eg.~\citep{tegmarkz02}). 

In this paper we vary four cosmological parameters, using flat priors on 
the baryon density $\Omega_{\rm b} h^2$, the
cold dark matter density $\Omega_{\rm c} h^2$,
the Hubble constant $h = H_0/(100 {\rm km~s}^{-1} {\rm Mpc}^{-1})$,
and the redshift of reionization $\zre$ (we assume $6 < \zre < 30$). 
The true ionization history may be complicated, so 
we assume there exists an effective redshift $\zre$ at
which a single rapid reionization gives a good approximation to the
true CMB anisotropy. The temperature CMB anisotropy is fairly
insensitive to the details of reionization, and the effect on the
polarization is only on large scales where it is largely hidden by
cosmic variance and the current observational noise, so this assumption 
should not affect our results significantly. 
Throughout we assume that the universe is flat with a 
cosmological constant. 
We assume purely Gaussian adiabatic scalar perturbations and
ignore tensor modes in this paper. 
Since tensor mode perturbations decay on sub-horizon scales they only affect
the large scale CMB anisotropy.
Given our assumptions, if we find excess power on large scales
this could, equivalently, be due to a tensor contribution rather than
the shape of the scalar initial power spectrum. 

We use the latest WMAP\footnote{\Url{http://lambda.gsfc.nasa.gov/}
}~\citep{Verde03,Hinshaw03,Kogut03}
temperature and temperature-polarization cross-correlation anisotropy
data.
We also include almost independent bandpowers on smaller
linear scales ($800< \ell < 2000$) from 
ACBAR\footnote{\Url{
http://cosmology.berkeley.edu/group/swlh/acbar/data
}}~\citep{Kuo02},
CBI~\citep{Pearson02} and the VSA~\citep{Grainge02}.

We constrain the matter power spectrum at low redshift by
using the galaxy power spectrum measurements
of the 2dFGRS~\citep{percivalea02}
over the range $0.02 <  k/(h \Mpci) < 0.15$. 
We assume that the galaxy power spectrum 
measured by the 2dFGRS
is a simple unknown multiple of the underlying matter power spectrum
(linear bias), so in our analysis the 2dFGRS data
serves to constrain the shape but not (directly) the amplitude of the matter
power spectrum.
In principle the matter power spectrum is directly proportional to the primordial
power spectrum
however in practice, due to the finite volume observed,
the data constrain a smeared version of the underlying 
matter power spectrum. This makes it harder to 
detect any sharp features in the primordial power spectrum,
as investigated by~\cite{elgaroygl02}.

To evaluate the posterior distributions of parameters given the data we
use the Markov-Chain Monte Carlo method to generate 
a list of samples  (coordinates in parameter space) 
such that the number density of samples is proportional
to the probability density. We use a modified version of the
CosmoMC\footnote{\Url{http://cosmologist.info/cosmomc/}}
code, using CAMB~\citep{lewiscl00} (based on CMBFAST~\citep{cmbfast}) to generate the CMB
and matter power spectrum transfer functions. 
CosmoMC uses the Metropolis-Hasting algorithm
to explore the posterior probability distribution in a piece-wise
manner, allowing us to exploit the fact that for each transfer
function computed many different values of the initial power spectra
parameters can be changed at almost no additional computational cost.
For further details see~\citet{cosmomc,christensenmkl01,Verde03}
and references therein.
Most of the chains for the analysis here were computed in around a day
using spare nodes of the CITA beowulf cluster, with each node running one
chain parallelized over the two processors. 
Between 4 and 20 converged chains
were generated for each set of parameters, burn in samples were
removed, leaving of the order of $10^5$--$10^6$ 
accepted positions in parameter space
from which the results in this paper were computed. 

\section{Power law spectra with and without a running spectral index}
\label{vanilla}

\begin{figure}
\centerline{\psfig{file = 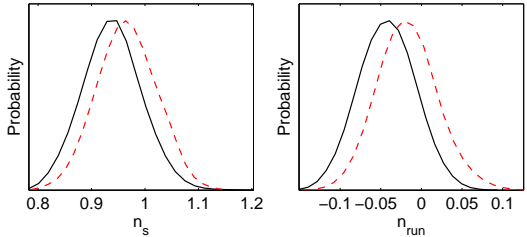,width=7cm}}
\caption{Marginalized distributions of the running spectral index
  slope parameters
  including (solid) and without (dashed) the WMAP temperature data at $\ell<5$.
}

\label{fig:nrun}
\end{figure}

In most parameter studies the scalar initial power spectrum is parameterized by a
constant
spectral index $n_s$ and an amplitude. Even with only two parameters defining the
primordial power spectrum there are large degeneracies between these
and the other cosmological parameters.
For example larger baryon densities decrease the height of the 
CMB second acoustic peak thereby mimicking the effect of a red 
spectral tilt.

Using the WMAP~\citep{Hinshaw03,Kogut03} power spectrum results alone 
we find a tight marginalized constraint 
$n_s - 30.4(\Omega_b h^2 - 0.025) = 1.04 \pm 0.02$ (68\% confidence).
However the orthogonal direction is very poorly
constrained with\footnote{This wide spread may be partly due the to
  approximations used in the WMAP likelihood parameterization.}
$n_s + 30.4(\Omega_b h^2 - 0.025) = 1 \pm 0.14$.
The best fit model to the WMAP data~\citep{Spergel03} with
$n_s = 0.97$ can be closely approximated by completely different
models with $n_s > 1.1$, higher reionization redshifts, rapid Hubble
expansion and high power spectrum amplitude. 
On integrating out the value of $\Omega_b h^2$ the constraint 
on the spectral index weakens to $n_s = 1.05 \pm 0.08$.
Similarly the amplitude of the primordial curvature fluctuations on 
$0.05{\rm Mpc}^{-1}$ scales is constrained by WMAP to be 
$A_s^{1/2} = (5.5\pm 0.8) \times 10^{-5}$, whereas the parameter combination 
$A_s^{1/2} e^{-\tau} = (4.3 \pm 0.1) \times 10^{-5}$
is much more tightly constrained
since it comes directly from the observed temperature anisotropy 
amplitude.
To partially break these degeneracies the WMAP team adopt a 
prior on the reionization optical depth of $\tau<0.3$.
By adding the 2dFGRS and additional CMB data at $\ell>800$ we find
the parameter constraints which broadly agree with the analysis reported by the WMAP 
team~\citep{Spergel03}, 
see Fig.~\ref{fig:pinibins_cospars} below.

\label{nrun}
By adding a running spectral index we find the marginalized 1-sigma result
$n_{\rm run}=-0.04 \pm   0.03$ 
shown by the solid line in the right hand panel of Fig.~\ref{fig:nrun}, 
in rough agreement with the WMAP team. 
For the pivot scale used ($k_{s0}=0.05 \Mpci$) a red tilt is preferred
(left hand panel of Fig.~\ref{fig:nrun}).
As expected, the effect of adding $n_{\rm run}$ as a free parameter
is to increase the uncertainties on the cosmological
parameters but only within the original uncertainties.

We find that the evidence for running comes predominantly from the very
largest scale multipoles. When we exclude $\ell<5$ 
multipoles from the WMAP temperature (TT) likelihood
the running spectral index distribution shifts to becomes highly 
consistent with $\nrun = 0$, as shown in  Fig.~\ref{fig:nrun} . 
The constraints on all cosmological parameters except for $n_{\rm s}$ and 
$n_{\rm run}$ are changed very little on removing the lowest multipoles.
The running
parameterization is therefore not ideally
suited to the data: WMAP
provides some evidence for low power on the very largest scales, but 
this is only
crudely fit by using a running spectral index. The WMAP analysis
relies on Lyman-$\alpha$ forest at wavenumbers greater than $k\sim0.1
{\rm Mpc}^{-1}$ to give evidence for more red tilt on small scales
consistent with a running index, but the validity of this analysis is
in serious doubt~\cite{Seljak03}.
We conclude that the marginal preference for a running spectral
index from our CMB + 2dFGRS analysis is primarily driven by the first
three CMB multipoles. 

\section{Power spectrum on the largest observable scales}
\label{cutoff}

\begin{figure}
\centerline{\epsfig{file=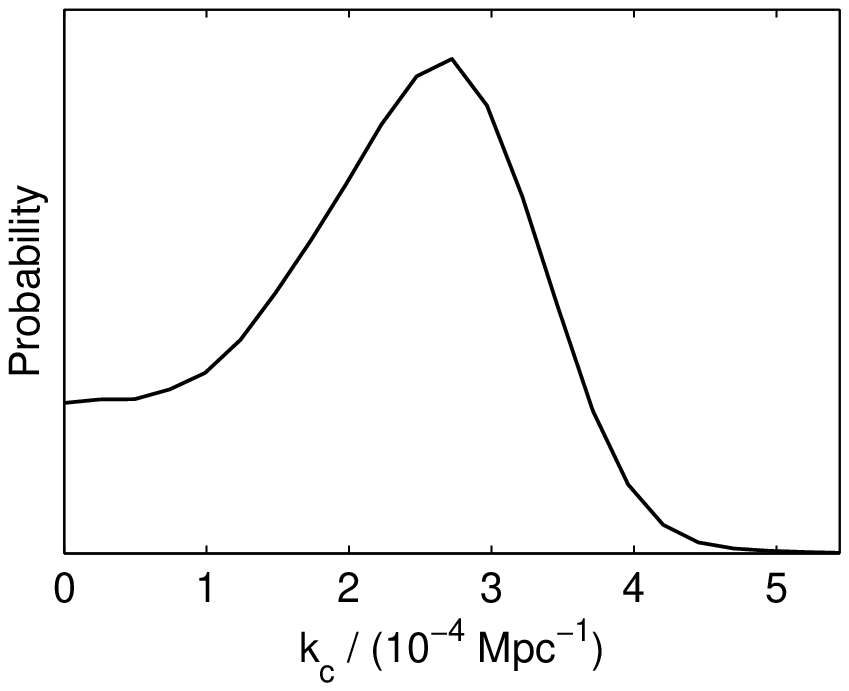, width=7cm}}
\centerline{\psfig{file = 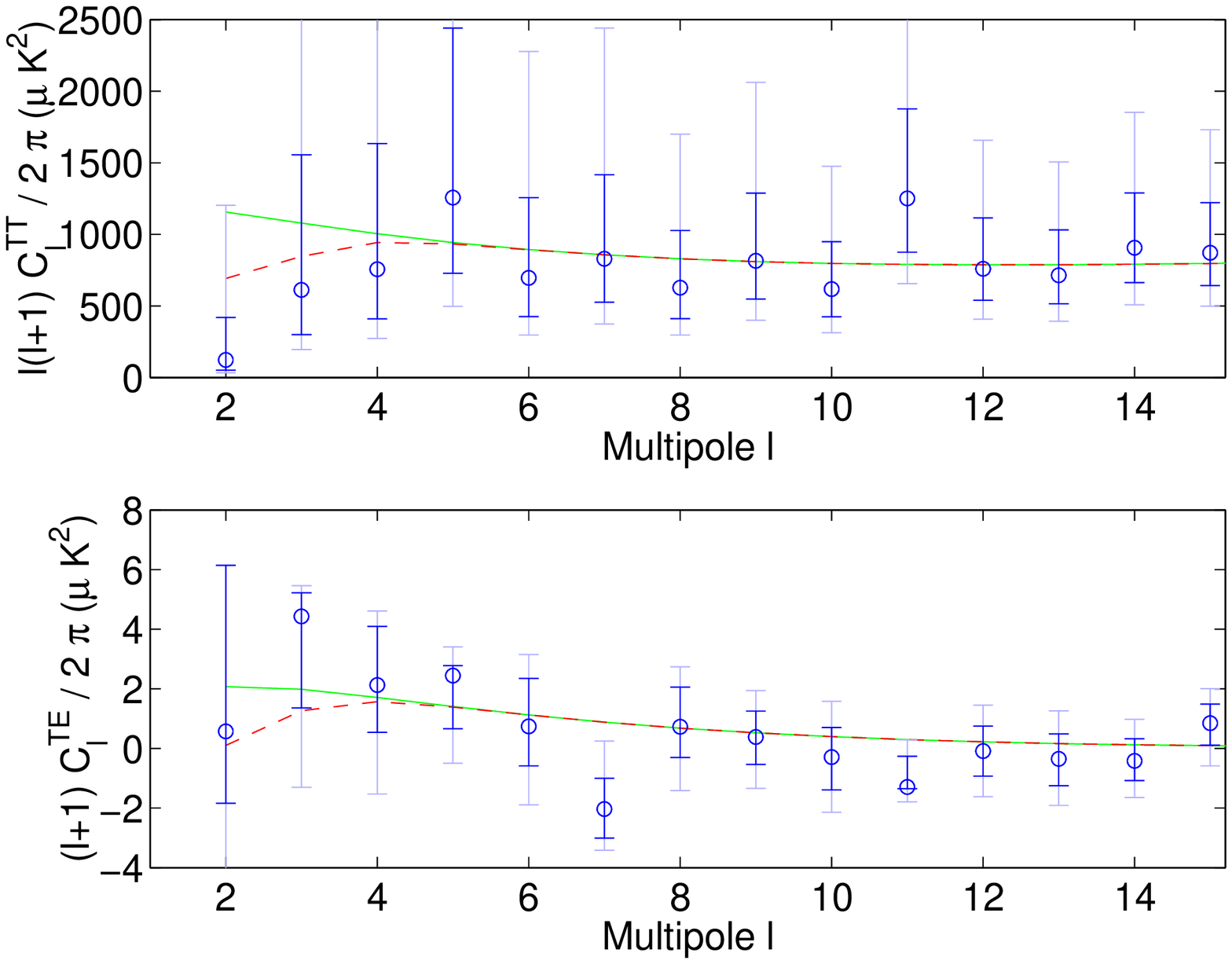, width=8.5cm}}
\caption{Top: Marginalized probability
distribution of a large scale power cut-off
  parameter $k_c$ for which ${\cal P}_\chi(k<k_c) = 0$. 
Lower panels: a concordance model with $k_c=0$ (solid lines) and with 
$k_c = 2.7\times 10^{-4}$ (dashed lines)
compared to the WMAP data.
Dark and light error bars show the 68 and 95 confidence limits 
on the theoretical value at each multipole (for the TE error
bars 
we use the Gaussian contribution to the likelihood given the observed TT estimator,
taking into 
account the correlation between the TT and TE estimators, assuming 
the WMAP best fit model for $C_{\ell}^{\rm TT}$ and $C_{\ell}^{\rm
  EE}$, where $(C_{\ell}^{\rm TE})^2<C_{\ell}^{\rm TT} C_{\ell}^{\rm EE}$).
}
\label{fig:cut}
\end{figure}

The quadrupole and octopole estimators observed by WMAP are low compared to the
other large scale multipoles. In a given model the low multipole estimators have a wide
$\chi^2$-like distribution (which has the peak below the
ensemble mean), so the low values could just be chance. However any
model that predicts small values for the low multipoles would be
favoured by the data, by a factor of up to about fourty, so this could
be a hint that there is a sharp fall in power on the largest scales.
\cite{Tegmark03} find that the quadrupole
and octopole appear to be aligned, perhaps indicative of some anisotropic
effect on 
on very large spatial scales which would not be well modeled by a
statistically isotropic power spectrum model. Here we assume that the
alignment is a coincidence, and proceed to consider whether the shape of the
initial power spectrum could help to explain the low large scale signal.

There is only a limited amount of information on the largest scales
due to cosmic variance, so we cannot hope to fit many extra parameters. 
We choose to assume there is a sharp total cut-off in
the primordial power on scales larger than $k=k_c$ 
(for a discussion of an exponential cut-off in closed models
see~\cite{Efstathiou03}; see~\cite{Contaldi03} for possible motivation for a cut-off). As discussed in the 
previous section a constant
spectral index is a good fit on smaller scales, so we parameterize the
primordial spectrum as 
\begin{equation}
\Pchi = \left\{
\begin{array}{c@{\qquad}l}
0 & k< k_c \\
A_s \left(\frac{k}{k_{s0}}\right)^{n_s-1}& k\ge k_c 
\,\,\,. 
\end{array}
\right.
\end{equation}
Marginalizing over the other parameters we find the constraint
on $k_c$ shown in the upper panel of
Fig.~\ref{fig:cut}.  
We find a preference for a cut-off at 
$k_c = (
2.7^{+0.5}_{-1.6} )\times 10^{-4} \Mpci$, a scale which can give a
significantly lower quadrupole and octopole without significantly
affecting the higher multipoles (lower panels of Fig.~\ref{fig:cut}).
However, according to this model a spectrum with no cut-off is not
strongly excluded by the data.
Our parameterization cannot achieve values for the quadrupole as low as
observed because there is a significant Integrated-Sachs Wolfe contribution
to the quadrupole from scales smaller than the cut.
The best-fit cut-off model has a very similar probability
to the best-fit running model, although the cut-off
model is marginally preferred. The constraints on the cosmological parameters are virtually 
unaffected by adding this free cut-off scale.

A comparison with~\cite{Spergel03}'s assessment of the significance of the low
CMB power at small multipoles is not straightforward because the
statistical tests differ.
A more detailed analysis of this important issue is required, but given
the results of Fig.~2 it would seem premature to discard simple continuous
power law models.

\section{Power spectrum reconstruction in wavenumber bands}
\label{pinibins}

On smaller scales there is now a wealth of data available to constrain
the primordial power spectrum in considerable generality. Here we
choose to parameterize the spectrum
in a number of bands, which has the advantage that it
is free from any assumptions about the underlying model. 
This generality can in principle reveal
unexpected features in the primordial power spectrum
on scales comparable to 
(or larger than) the band width. The disadvantage is that there
are a large number of free parameters so care is needed to obtain
meaningful results.

The initial power spectrum is only indirectly constrained by the
observational data. Since we do not model the biasing in the 2dFGRS
galaxy power spectrum, the only direct constraint on the amplitude 
(as opposed to the shape) comes from the CMB anisotropy power spectra.
On scales smaller than the horizon size at reionization
($\ell \sim 40$) the observed
power in the CMB anisotropy scales as $e^{-2\tau} \Pchi$, 
depending on the optical depth to reionization $\tau$.
We therefore parameterize the shape of the initial power
spectrum by the value of $e^{-2\tau} \Pchi$ at a set of points, and 
linearly interpolate 
(in $k$) between the points. We then refer to the power at
each point as the power in a `band' at
that wavenumber.
Our parameters $b_i$ are defined by 
\begin{equation}
e^{-2\tau} \Pchi = \left\{
\begin{array}{c@{\qquad}l}
\frac{ (k_{i+1}-k)b_i + (k-k_i) b_{i+1}}{k_{i+1}-k_i}
& k_i < k < k_{i+1} \\
b_n & k>k_n
\end{array}
\right.
\label{eq:pinibins}
\end{equation}
where the first line applies for $1 \le i \le n-1$.
We set the position of the first band to $k_1=0$
and that of the second band to $k_2=0.005 {\rm Mpc}^{-1}$, so we only
have two bands
over the region where the data are effected by the modes
which are super-horizon at reionization and hence do not scale simply
with $e^{-2\tau}$.
Subsequent band positions on smaller scales are logarithmically spaced with
$k_{i+1}=1.275 \, k_{i}$,
where the constant is chosen such that $k_{16}=0.15 {\rm
  Mpc}^{-1}$. We assume a flat prior on each of the $b_i$. This prior is equivalent
to a flat prior on the underlying power spectrum 
$e^{2\tau}
b_i$, together with a prior on the optical depth $P(\tau) \propto e^{-2n\tau}$.
Constraints on quantities
sensitive to $\tau$ 
depend strongly on the choice
prior for large numbers of bands. This is one reason we reconstruct $b_i$, which are
more directly constrained by the data and have the main dependence on
$\tau$ taken out, rather than $\Pchi$ directly. Our choice of prior gives a posterior constraint on
the optical depth similar to (but somewhat broader than)  that with the simple spectral index parameterization.

\begin{figure}
\epsfig{ file=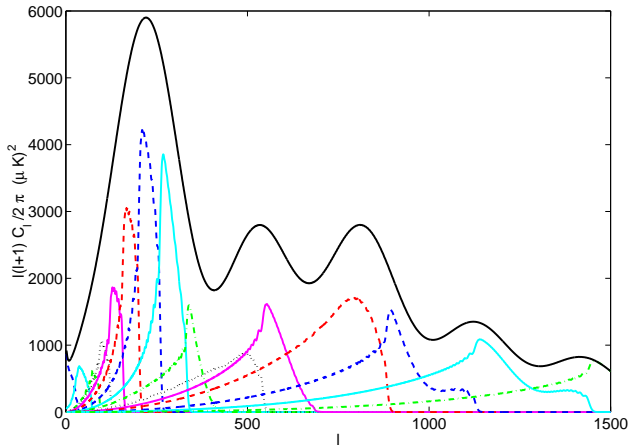, width=6cm, angle = -90}
\caption{CMB power spectra for top hat primordial power
spectra at the positions of the bins used for Fig.~\ref{fig:pinibins}.
The top line is the sum of all the lower lines, ie. the usual
CMB power spectrum.
}
\label{fig:pinibins_ktoell}
\end{figure}

\begin{figure*}
\epsfig{ file=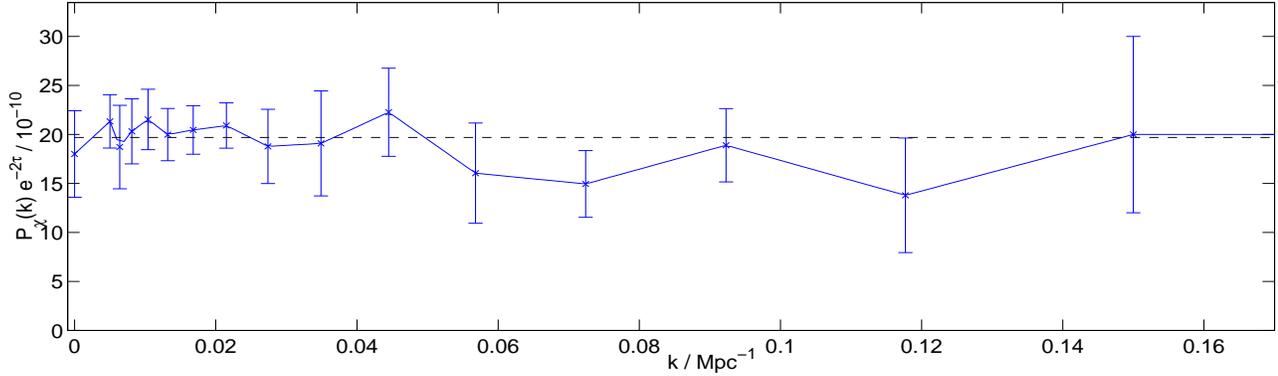, height=5cm, width=17cm}
\caption{Reconstruction of the shape of the primordial power spectrum
in 16 bands after marginalising over
the Hubble constant,  baryon and dark matter densities, and
the redshift of reionization. }
\label{fig:pinibins}
\end{figure*}

Fig.~\ref{fig:pinibins_ktoell} shows the CMB temperature power spectra for 
top hat primordial power spectra which are zero except for
the range $(k_{i-1}+k_i)/2<k<(k_i+k_{i+1})/2$ 
(except for $i=1$ where $0<k<k_2/2$) for a concordance model. 
All the primordial power spectrum
top hats have the same absolute height.
The result illustrates the rough correspondence between the position
of the bands we use and the CMB power spectrum peaks, showing that there is one
band covering much of the large scales, and several bands over the
first acoustic peak, with only one or two bands for 
each of the subsequent few peaks.

Our reconstruction of the shape of the initial power spectrum is shown in Fig.~\ref{fig:pinibins}.
The crosses and error bars show the 
peak and 
68\% upper and lower limits of the marginalized distributions.
The probability distribution of 
all but the last band is close to Gaussian,
 although some bands are correlated at
around the fifty per cent level. 
The maximum correlation is 80 per cent, which is a positive correlation 
between bands around the first acoustic peak of the CMB.
We have checked that our constraints on the sub-horizon $b_i$ agree well with those obtained by
fixing the optical depth to $\tau = 0.17$ rather than marginalizing
over it.
The overall shape is consistent with a featureless
scale invariant power spectrum, though
perhaps an overall red tilt is
discernible,
bearing in mind the large error bar on the small scale point.
We do not see the feature at $k\sim0.015 {\rm Mpc}^{-1}$ 
suggested by~\cite{mukherjeew03a}.
Our reconstruction is complementary to that in~\cite{mukherjeew03c}
since our inter-band separation is one third of theirs.
Also, we use linear interpolation instead of wavelets or top-hat bins,
which makes the primordial power spectrum relatively smooth.

\begin{figure}
\epsfig{ file=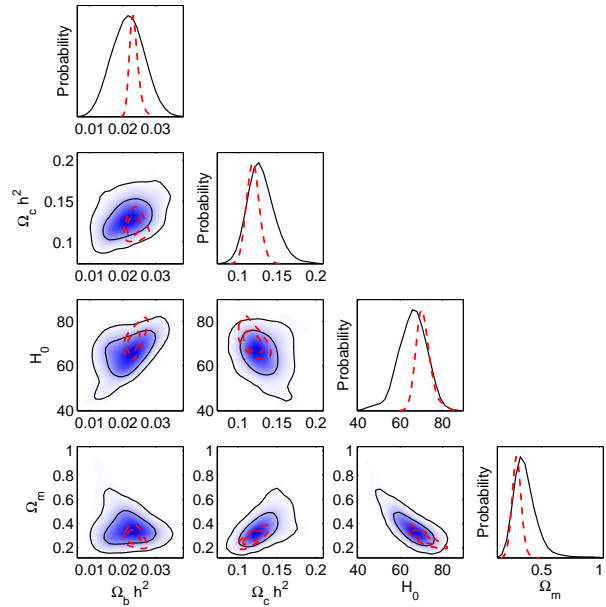, width=8cm}
\caption{
Constraints on cosmological parameter estimates 
after marginalising over the power spectrum amplitudes in 16 bands
(solid lines), and for comparison equivalent results in the  
constant
spectral index model (dashed lines). Contours enclose $68\%$ and $95\%$ of the probability.}
\label{fig:pinibins_cospars}
\end{figure}

In Fig.~\ref{fig:pinibins_cospars} we investigate how robust the
cosmological parameter estimates are to this general primordial
power spectrum.
Dashed lines show the cosmological parameter 
constraints assuming a power law primordial power 
spectrum, marginalising over $n_s$. 
Overlaid, the solid lines show the result after allowing
freedom in the amplitudes of the 16 bands.
The error contours are broader but remarkably show that
it is still possible to recover interesting constraints 
on the cosmological parameters
even if one allows the large amount of freedom in the 
primordial power spectrum shape.

It is important to emphasize that this method of power spectrum
differs from earlier work such 
as that by~\cite{gawisers98} because the
MCMC reconstruction properly accounts for the allowed ranges
of all of the parameters defining the model.

\section{Broken power spectrum}
\label{barriga}

\begin{figure}
\epsfig{ file=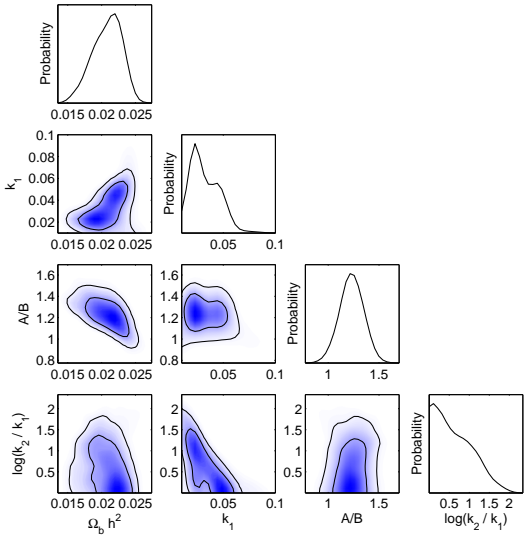, width=7cm}
\caption{Marginalized constraints on the parameters of the broken
  power spectrum model and the  correlations with the baryon density.
The constraint on $\tau$ is changed relatively little on allowing the 
16 bands, becoming $\tau=0.13 \pm 0.07$
}
\label{fig:bar}
\end{figure}

\begin{figure}
\epsfig{ file=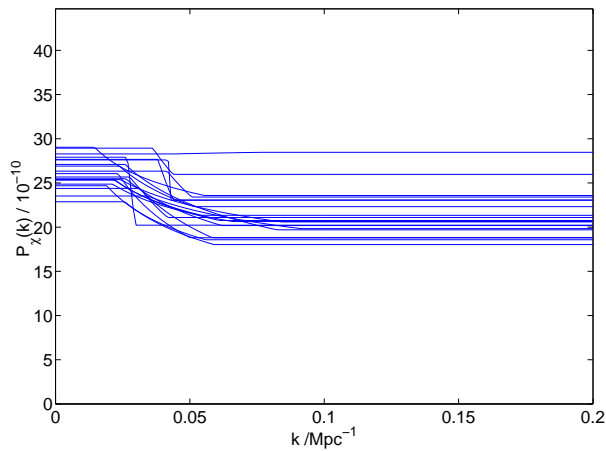, width=8cm}
\caption{Primordial power spectra of
random samples from the broken power spectra chains.}
\label{fig:bar_pk}
\end{figure}

In this section we use a parameterization motivated by double
inflation, previously explored by ~\cite{barrigagss01}.
In this model
\begin{equation}
{\cal P}_{\chi} (k) = \left\{
\begin{array}{c@{\qquad}l}
 A & k<k_1 \\
 C k^{\alpha-1} & k_1<k<k_2 \\
B & k>k_2 \,\, .
\end{array}
\right.
\end{equation}
where $C$ and $\alpha$ are chosen such that the power spectrum is continuous.
As in~\citet{barrigagss01} we take flat priors on 
$k_1$, $A/B$ and   $\log ( k_2/k_1)$ and we limit 
the parameter space to $0.01<k_1 / {\rm Mpc^{-1}}<0.1$, $0.3<A/B<7.2$ and 
$0.01 < \log (k_2/k_1) <4 $. 
In addition we impose a prior $k_2<0.1 {\rm Mpc^{-1}}$ so that
we explore a transition only in the region probed
by the observational data.

In Fig.~\ref{fig:bar} we show the constraints on 
the model parameters after marginalising over
four cosmological parameters.
A conventional scale-invariant spectrum corresponds
to $A/B=1$ and we can see that this possibility
is very close to the 1$\sigma$ contours.
Values of $A/B$ higher than unity are preferred,
corresponding to a drop in the initial power spectrum
on going from large to small scales.
This is not surprising since the data favour red tilts, and in this
parameterization a tilt is obtained by having a long transition
between a higher flat spectrum on large scales and a lower flat
spectrum on small scales.
The distribution is slightly bimodal, with the first mode
roughly corresponding to a wide transition straddling the
first CMB acoustic peak and the second corresponding to a 
drop in power in the dip between the first and second peaks.
The expected strong correlation between $A/B$ and 
the baryon density is clear.
We find that removing the first three multipoles has no effect
on the parameter constraints on this model.

We show in Fig.~\ref{fig:bar_pk} the power
spectra corresponding to 20 samples in the MCMC chain.
There is considerable spread, but the transition,
if any, occurs at around $k \sim 0.04 {\rm Mpc^{-1}}$.
This corresponds roughly to the scale at which the 2dFGRS
data becomes statistically significant.
Since the 2dFGRS data may disfavour a transition
at higher wavenumber and the WMAP data would disfavour
a transition at lower wavenumber then neither the 
amplitude or the scale of the break is likely to be 
of any physical significance.

The effect of adding these additional parameters on the estimates
of cosmological parameters is small; the biggest effect is to
widen the allowed range of $\Omega_{\rm b} h^2$ to include smaller values.
This is due to the fact that a 
preferred position for any transition is between 
the first two acoustic CMB peaks, and it is also
this ratio that determines the baryon density.

\section{Conclusions}
\label{conclusions}

We have explored various parameterizations of the primordial 
power spectrum and found that in each case a 
simple scale invariant spectrum is an acceptable fit to the data.
Some deviation towards a red tilt is preferred and there is
a marginal indication of
a cut-off in power on the largest observable scales. Significantly, 
even if the power spectrum is allowed to vary
over a large number of wavenumber bands on sub-horizon scales the reconstructed
spectrum is found to be featureless and close to scale invariant.
In addition we have shown how the constraints on the other
cosmological parameters are affected by the addition
of this extra freedom in the primordial power spectrum shape:
the error bars on cosmological parameters
are significantly amplified but the resulting constraints
are still strong and roughly similar to those
before the recent WMAP results.
The unexpectedly small power on superhorizon
scales observed by WMAP, if real, provides marginal evidence for
a drop in the primordial power spectrum on the largest observable scales.

\section*{Acknowledgments}

We thank Carlo Contaldi and the Cambridge Leverhulme Group for helpful discussions,
in particular Anze Slosar, Ofer Lahav, Anthony Lasenby, 
Dan Mortlock, Carolina Odman and Mike Hobson.
SLB acknowledges support from PPARC and a Selwyn College Trevelyan Research
Fellowship.
JW is supported by the Leverhulme Trust and Kings College
Trapnell Fellowship.
The beowulf computer used for this analysis was funded by the Canada 
Foundation for Innovation and the Ontario Innovation Trust. 
We also used
the UK National Cosmology Supercomputer Center funded by
PPARC, HEFCE and Silicon Graphics / Cray Research.

\label{lastpage}

\end{document}